\documentclass[aps,pra,twocolumn,longbibliography]{revtex4-2}
\usepackage{amsmath}
\usepackage{epsfig}
\usepackage{epstopdf}
\usepackage{graphicx}
\usepackage{rotating}
\usepackage{amssymb}
\usepackage{float}
\usepackage[final,inline,nomargin]{fixme} \fxsetup{theme=color}
\usepackage{braket}
\usepackage{subfigure}
\usepackage{natbib}
\usepackage{bm}
\usepackage[usenames,dvipsnames]{xcolor}
\usepackage[colorlinks=true,citecolor=Cerulean,linkcolor=RubineRed,urlcolor=Cerulean]{hyperref}

\fxsetup{status=draft}

\begin{document}

\title{Estimation of the condensate fraction from the static structure factor}
\author{Yu. E. Lozovik$^1$, I. L. Kurbakov$^1$, G. E. Astrakharchik$^2$, and J. Boronat$^2$}
\affiliation{$^1$Institute of Spectroscopy (Russian Academy of Sciences), 108840, Troitsk, Moscow, Russia}
\affiliation{$^2$Departament de F\'{\i}sica, Campus Nord B4-B5, Universitat Polit\`ecnica de Catalunya, E-08034 Barcelona, Spain}
\date{April 7, 2021}

\begin{abstract}
We present an analytical method to estimate the condensate fraction $n_0/n$ in strongly correlated systems for which the zero-temperature static structure factor $S({\bf p})$ is known. The advantage of the proposed method is that it allows one to predict the long-range behavior of the one-body density matrix (i) in macroscopic and mesoscopic systems, (ii) in three- and two-dimensional geometry, (iii) at zero and low finite temperature, and (iv) in weakly and strongly correlated regimes. Our method is tested against exact values obtained with various quantum Monte Carlo methods in a number of strongly correlated systems showing an excellent agreement. The proposed technique is also useful in numerical simulations as it allows one to extrapolate the condensate fraction to the thermodynamic limit for particle numbers as small as tens to hundreds. Our method is especially valuable for extracting the condensate fraction from the experimentally measured static structure factor $S({\bf p})$, thus providing a new simple alternative technique for the estimation of $n_0/n$. We analyze available experimental data for $S({\bf p})$ of superfluid helium and find an excellent agreement with the experimental value of $n_0/n$.
\end{abstract}
\maketitle

\section{Introduction}
Strongly-interacting Bose systems, such as excitons in quantum wells~\cite{na0483000584, epl107010012, pnas11618328} and transition metal dichalcogenides (TMDs)~\cite{na0499000419, nc0005004555}, ultracold gases\cite{prx009021012} in optical lattices~\cite{sci352000201}, at Feshbach resonance~\cite{Jin2014, Hadzibabic2017, ChevySalomon2016}, Rydberg atoms~\cite{prl114203002, np0012001095} as well as superfluid $^4$He~\cite{glyden0}, are now attracting a special attention. Whereas weakly correlated systems are routinely described by mean-field theory~\cite{prb053009341}, the regime of strong correlations hardly allows for an analytical description. Importantly, the Bose-Einstein condensate (BEC) density, being the main magnitude quantifying the macroscopic coherence, cannot be calculated perturbatively in the regime of strong correlations. This failure is caused by the large condensate depletion produced by interparticle correlations. For example, the condensate fraction of superfluid $^4$He is only 7\%\cite{glydebook} and in two-dimensional dipolar excitons in GaAs quantum wells~\cite{prl098060405,ssc144000399} is 40\% \cite{na0483000584}, 25\% \cite{nl0007001349} or even smaller, 10\% \cite{jetpl0840222}. Large condensate depletion entails significant complications for a precise determination of the condensate density both in experiment and in theory. Unfortunately, at present there is not a simple analytical theory capable of making a quantitative prediction for the condensate fraction in the regime of strong correlations. The experimental estimation of the condensate fraction in liquid $^4$He was done, among others, by Glyde and collaborators~\cite{glydet}, after intensive neutron scattering experiments. The analysis of the experiment is quite elaborated as final state effects (FSE) in the scattering process have to be taken into account. The estimation of the FSE is rather involved and several theories have been devised to account for them, normally acting as a convolution function applied to the impulse approximation~\cite{gersch1,gersch2,Silver1,Silver2,Silver3,Rinat1,Rinat2,Carraro1,Carraro2,glyde,mazzanti}. The experimental values obtained in this way are in close agreement with quantum Monte Carlo (QMC) simulations in a wide range of densities~\cite{glyden0}. Still, high-precision neutron scattering experiments are quite complicated and expensive so this field would benefit from a simpler alternative to estimate the condensate fraction.

In this article, we establish a new useful relation between the condensate fraction $n_0/n$ and the zero-temperature static structure factor $S({\bf p})$ with $n$ ($n_0$) being the total (condensate) density. We present a new, fully analytical method for describing the long-wavelength properties of cold bosonic systems in terms of $S({\bf p})$, which enters as the only input quantity. Our method is based on an empirical choice of the ultraviolet (UV) cutoff in quantum-field hydrodynamics (HD)\cite{pr0155000080,Popov,prb049001205,prb049012938}. We exploit a number of advanced quantum Monte Carlo (QMC) methods (variational, diffusion, path-integral) in order to verify our theory. We find a remarkable quantitative agreement in a number of mesoscopic and macroscopic systems and in different dimensionalities (2D and 3D), both at zero and low temperatures. Our method is valid, not only in the Bogoliubov perturbative regime, but in the regime of strong correlations, since the small parameter in our theory is the normal fraction $(n-n_s)/n$ rather than the non-condensate fraction $(n-n_0)/n$, with $n_s$ the superfluid density. Remarkably, and in contrast to standard approaches\cite{pr0155000080, Popov, prb049001205, prb049012938}, we obtain a quantitative agreement with QMC results for a spatial range as small as few interparticle distances. This allows one to make a reliable extrapolation to the thermodynamic limit based on simulations performed only with hundreds or even tens of particles. As a further check, we apply our method to probably the most famous and difficult example of a strongly correlated system, namely superfluid Helium. Even if the condensate fraction is as small as 7\%, our method is able to reproduce it correctly when QMC data for $S({\bf p})$ is taken as an input. Finally, we use experimental data\cite{glyden0, wirth, svensson} of $S({\bf p})$ and find that the condensate fraction obtained in this way are in agreement with the ones derived from deep-inelastic neutron scattering. This opens the door to an alternative and easier way of determining the condensate fraction of quantum systems in which strong correlations produce a deep depletion of the condensate.

\section{Hydrodynamic theory}\label{HDtheory}
We consider a bulk homogeneous three-dimensional (3D) or two-dimensional (2D) bosonic superfluid system in absence of static currents. The effective Hamiltonian corresponding to the free-energy functional can be conveniently written in the form~\cite{Popov,prb049012938}
\begin{equation}\label{H}
\hat H-\mu\hat N=\int\left(\frac m2\hat\rho_s({\bf r})
(\hat{\bf v}'({\bf r}))^2+ f(\hat\rho'({\bf r}))\right)d{\bf r} \ ,
\end{equation}
with $m$ the particle mass, $\mu$ the chemical potential, and $\hat N$ the particle number operator. The density operator $\hat\rho_s({\bf r})$ of the superfluid component is conveniently split into the superfluid density $n_s$ and the density fluctuation operator $\hat\rho'({\bf r})$ as $\hat\rho_s({\bf r}) = n_s + \hat\rho'({\bf r})$. The density $\hat\rho'({\bf r})$ and velocity $\hat{\bf v}'({\bf r})$ fluctuation operators of the superfluid component have zero Gibbs averages, $\langle\hat\rho'({\bf r})\rangle = \langle\hat{\bf v}'({\bf r})\rangle = 0$ and zero volume averages, $\int\hat\rho'({\bf r})d{\bf r}=\int\hat{\bf v}'({\bf r})d{\bf r}=0$. As usual, brackets $\langle \cdots \rangle$ denote averaging over the equilibrium state of the system that preserves the number of particles. Finally, the function $f(\hat\rho'({\bf r}))$ in Eq.~(\ref{H}) describes two-, three- or many-body interactions.

The Hamiltonian~(\ref{H}) $\hat H-\mu\hat N$ can be decomposed into quadratic and anharmonic terms. In order to find the correlation functions, we need only the quadratic form
\begin{equation}\label{H0r}
\hat H_0=\int\left(\frac{mn_s}2(\hat{\bf v}'({\bf r}))^2+
\frac g2(\hat\rho'({\bf r}))^2\right)d{\bf r} \ ,
\end{equation}
with $g\equiv m^2 / \chi$ a generalization of the coupling constant to the case of strong correlations and $\chi$ being the adiabatic compressibility~\cite{prb049012938}. Let us expand the density and velocity fluctuation operators in Fourier series,
\begin{equation}\label{rhovr}
\hat\rho'({\bf r})=\frac1{\sqrt{V_D}}\sum_{{\bf p}\ne0}
e^{i{\bf pr}/\hbar}\hat\rho_{\bf p},\;\;
\hat{\bf v}'({\bf r})=\frac1{\sqrt{V_D}}\sum_{{\bf p}\ne0}
e^{i{\bf pr}/\hbar}\hat{\bf v}_{\bf p},
\end{equation}
where $D=2$ or 3 is the system dimensionality and $V_D = N/n$ is the quantization volume. We decompose the velocity fluctuation operator $\hat{\bf v}_{\bf p}$ into phononic $\hat{\bf v}_{\bf p}^{\parallel}$ and vortex $\hat{\bf v}_{\bf p}^{\perp}$ contributions, where $\hat{\bf v}_{\bf p}^{\parallel}\parallel{\bf p}$ and $\hat{\bf v}_{\bf p}^{\perp}\perp{\bf p}$, with $\hat{\bf v}_{\bf p}^{\parallel}+\hat{\bf v}_{\bf p}^{\perp}=\hat{\bf v}_{\bf p}$. Then, the quadratic Hamiltonian~(\ref{H0r}) in reciprocal space turns to
\begin{equation}\label{H0p-total}
\hat H_0=\hat H_0^{\parallel}+\sum_{{\bf p}\ne0}
\frac{mn_s}2\hat{\bf v}_{\bf p}^{\perp}\hat{\bf v}_{\bf -p}^{\perp},
\end{equation}
where
\begin{equation}\label{H0p}
\hat H_0^{\parallel}=\sum_{{\bf p}\ne0}
\left(\frac{mn_s}2\hat{\bf v}_{\bf p}^{\parallel}\hat{\bf v}_{\bf -p}^{\parallel}+
\frac g2\hat\rho_{\bf p}\hat\rho_{\bf -p}\right)
\end{equation}
is the phononic part of the Hamiltonian.

Following Popov~\cite{Popov}, we generalize the standard hydrodynamic Hamiltonian~(\ref{H0p}) by introducing the momentum dependence of the coupling constant $g$,
\begin{equation}\label{g-->gp}
g\longrightarrow g({\bf p})\equiv\frac{p^2}{4mn_s}+U({\bf p})\ .
\end{equation}
Here, $U({\bf p})$ is a momentum-dependent potential satisfying the limiting 
condition $U({\bf p}\to 0)=g$. After 
substituting Eq.~(\ref{g-->gp}) into Eq.~(\ref{H0p}) the hydrodynamic 
Hamiltonian takes the following form
\begin{equation}\label{H0Up}
\hat H_0^{\parallel}=\sum_{{\bf p}\ne0}
\left[\frac{mn_s}2\hat{\bf v}_{\bf p}^{\parallel}\hat{\bf v}_{\bf -p}^{\parallel}+
\left(\frac{U({\bf p})}2+\frac{p^2}{8mn_s}\right)
\hat\rho_{\bf p}\hat\rho_{\bf -p}\right] \ .
\end{equation}

The zero-temperature static structure factor $S({\bf p}) = \langle\hat\rho_{\bf p}\hat\rho_{\bf -p}\rangle/n$ at $T=0$ is our single input parameter. Therefore, a proper theory must reproduce the static structure factor self-consistently. Under the conditions of interest, $n-n_s\ll n$, this dictates the following choice for the momentum-dependent potential,
\begin{equation}\label{Up}
U({\bf p})=\frac{p^2}{4mn_s}\left(\frac1{(S({\bf p}))^2}-1\right)
\end{equation}
with the $U({\bf p}\to 0)=g$ condition being automatically held. Then, the empirical Hamiltonian~(\ref{H0Up}) is diagonalized by the transformation
\begin{equation}\label{rhopvp}
\hat\rho_{\bf p}=i\sqrt{n_sS({\bf p})}(\hat c_{\bf p}-\hat c_{\bf -p}^+),\;
\hat{\bf v}_{\bf p}^{\parallel}=\frac{i{\bf p}/m}{\sqrt{4n_sS({\bf p})}}
(\hat c_{\bf p}+\hat c_{\bf -p}^+),
\end{equation}
with ${\bf p}\ne0$. In Eqs.~(\ref{rhopvp}), $\hat c_{\bf p}$ is the annihilation operator of a phonon with momentum ${\bf p}$, satisfying the usual bosonic commutation relations~\cite{prb049001205},
$[\hat c_{\bf p},\hat c_{\bf q}]=0$, $[\hat c_{\bf p},\hat c_{\bf q}^+]=\delta_{\bf pq}$. 
We recast the empirical Hamiltonian~(\ref{H0Up}) in the final form
\begin{equation}\label{H0c}
\hat H_0^{\parallel}=\sum_{{\bf p}\ne0}\tilde\varepsilon_{\bf p}
\hat c_{\bf p}^+\hat c_{\bf p},\;\;\;\;\;\;
\tilde\varepsilon_{\bf p}\equiv\frac{p^2}{2mS({\bf p})} \ ,
\end{equation}
where the energy of the quasi-particles $\tilde\varepsilon_{\bf p}$ corresponds to the Feynman excitation energy. Therefore, $\tilde\varepsilon_{\bf p}$ is an empirical dispersion relation, which differs from the true one, and coincides with it only in the phononic long-wavelength limit $p\to 0$, where the described approach is valid.

Our main interest focuses on the estimation of the one-body density matrix. To this end, we use the Kubo cumulant expansion~\cite{pr0155000080},
\begin{equation}\label{obdm1}
g_1\!({\bf r})\!\equiv\!\langle\hat\Psi\!^+\!({\bf r})\!\hat\Psi(0)\rangle
\!=\!{\rm const} \times \exp\!\!\left[\!-\frac{\langle(\hat\varphi({\bf 
r})\!-\!\hat\varphi(0))^2\rangle}{2}\!\right],
\end{equation}
where $\hat\Psi({\bf r})$ is the operator of the Bose particle field, and the constant prefactor depends on the particular choice of the ultraviolet cutoff. The phase fluctuation operator $\hat\varphi({\bf r})$ can be expressed in terms of the superfluid velocity $\hat{\bf v}({\bf r})$, 
\begin{equation}\label{phir0}
\hat\varphi({\bf r})-\hat\varphi(0)\equiv\frac m{\hbar}\int_0^{\bf r}
\hat{\bf v}'({\bf r}')d{\bf r}' \ .
\end{equation}
We assume sufficiently low temperatures so that the thermal activation of the vortex rings (3D) or vortices (2D) is exponentially suppressed, and we set the corresponding contribution to zero, $\hat{\bf v}_{\bf p}^{\perp}=0$. In the 3D case, vortex effects are not important up to sufficiently high temperatures, where our method is no longer expected to be applicable. In the 2D case, the vortex effects must be taken into account, since they are necessary for an appropriate description of the BKT phase transition. Renormalization by free vortices (if they are generated) can be taken into account following Kosterlitz~\cite{jpc007001046} by introducing a $e^{-r/\xi_+}$ factor in Eq.~(\ref{obdm1}) with $\xi_+$  the distance between free vortices~\cite{prl040000783}. Instead, a renormalization originating from pairs of vortices can be included in $n_s$~\cite{prb024002526}. Therefore, in both 2D and 3D cases we arrive at the result
\begin{equation}\label{obdm2}
g_1({\bf r})={\rm const}\times \exp[-\langle(\hat\varphi_{\parallel}({\bf r})
-\hat\varphi_{\parallel}(0))^2/2\rangle]e^{-r/\xi_+} \ ,
\end{equation}
where the phononic contribution to the phase fluctuation is
\begin{equation}\label{phir0||}
\hat\varphi_{\parallel}({\bf r})-\hat\varphi_{\parallel}(0)=
\frac1{\sqrt{V_D}}\sum_{{\bf p}\ne0}\frac{m{\bf p}}{ip^2}
\hat{\bf v}_{\bf p}^{\parallel}(e^{i{\bf pr}/\hbar}-1).
\end{equation}
In two dimensions, if free vortices are absent (below the BKT critical temperature) one should set the corresponding distance $\xi_+$ to infinity. In three dimensions, one should always set $\xi_+=\infty$.

After substituting Eq.~(\ref{phir0||}) into Eq.~(\ref{obdm2}) with $\xi_+=\infty$ one has
\begin{equation}\label{obdm3}
\frac{g_1({\bf r})}{n}=\exp\left[\frac1{V_D}\sum_{{\bf p}\ne0}
\kappa_{\bf p}({\bf r})\frac{m^2}{p^2}
\langle\hat{\bf v}_{\bf p}^{\parallel}\hat{\bf v}_{\bf -p}^{\parallel}\rangle
\left(\cos\frac{\bf pr}{\hbar}-1\right)\right],
\end{equation}
where we inserted an ultraviolet cutoff factor $\kappa_{\bf p}({\bf r})$ ``by hand'' $ $ and take into account the fact that $g_1(0)=n$.

Averages of Bose operators satisfy $\langle\hat c_{\bf p}\rangle=\langle\hat c_{\bf p}\hat c_{\bf q}\rangle=0$ and
$\langle \hat c_{\bf p}^+\hat c_{\bf q}\rangle=\delta_{\bf pq}/(e^{\tilde\varepsilon_{\bf p}/T}-1)$, thus the velocity-velocity correlation function becomes [see Eq.~(\ref{rhopvp})]
\begin{equation}\label{<vv>}
\langle\hat{\bf v}_{\bf p}^{\parallel}\hat{\bf v}_{\bf -p}^{\parallel}\rangle=
\frac{p^2}{4m^2n_sS({\bf p})}
\frac{e^{\tilde\varepsilon_{\bf p}/T}+1}{e^{\tilde\varepsilon_{\bf p}/T}-1}.
\end{equation}
Within first-order perturbation theory, there is no distinction between the total and superfluid densities. Thus in Eq.~(\ref{<vv>}) we substitute $n_s$ by $n$ and, by using Eq.~(\ref{obdm3}), we arrive to the following final expression for the OBDM in a finite system,
\begin{equation}\label{obdm-final-meso}
\frac{g_1({\bf r})}{n}=\exp\left[\frac1{V_D}\sum_{{\bf p}\ne0}
\frac{\kappa_{\bf p}({\bf r})}{4nS({\bf p})}
\frac{e^{\tilde\varepsilon_{\bf p}/T}+1}{e^{\tilde\varepsilon_{\bf p}/T}-1}
\left(\cos\frac{\bf pr}{\hbar}-1\right)\right].
\end{equation}
In the thermodynamic limit, the summation over momenta in  Eq.~(\ref{obdm-final-meso}) should be replaced by an integral. This yields the final form for OBDM in a macroscopic system
\begin{equation}\label{obdm-final-macro}
\frac{g_1({\bf r})}{n}=\exp\left[\int\frac{d{\bf p}}{(2\pi\hbar)^D}
\frac{\kappa_{\bf p}({\bf r})}{4nS({\bf p})}
\frac{e^{\tilde\varepsilon_{\bf p}/T}+1}{e^{\tilde\varepsilon_{\bf p}/T}-1}
\left(\cos\frac{\bf pr}{\hbar}-1\right)\right].
\end{equation}
The integral is convergent in both the infrared and ultraviolet integration limits, $p \to 0$ and $p\to\infty$, if we take the cutoff factor in the form of 
\begin{equation}
\label{SM:kpr}
\kappa_{\bf p}({\bf r})=|1-S({\bf p})|^{2/(2-g_1^{T=0}({\bf r})/n)} \ ,
\end{equation}
and consider that the structure factor converges sufficiently fast to unity at large momenta. The cutoff function~(\ref{SM:kpr}) satisfies the conditions $\kappa_{{\bf p}\to0}({\bf r})=1$ and $\kappa_{{\bf p}\to\infty}({\bf r})=0$ at any distance $r$, as it must be in a proper HD theory~\cite{Popov}. As the OBDM also appears in the cutoff, the calculation of $g_1({\bf r})$ has to be done by solving an algebraic equation for $g_1^{T=0}({\bf r})$. The efficiency of our approach is based on the reduction of the complicated Hamiltonian~(\ref{H}) to a simple quadratic in $\hat\rho'$ and $\hat{\bf v}'$ form~(\ref{H0Up}).

Substituting Eq.~(\ref{SM:kpr}) into (\ref{obdm-final-macro}) and taking the $r\to\infty$ limit, we obtain the ground-state macroscopic condensate density $n_0^0=g_1^{T=0}(\infty)$ as a solution of the following algebraic equation
\begin{equation}\label{n0n-final-macro}
\frac{n_0^0}{n}=\exp\left[-\int\frac{d{\bf p}}{(2\pi\hbar)^D}
\frac{\displaystyle|1-S({\bf p})|^{2/(2-n_0^0/n)}}{4nS({\bf p})}\right]\;.
\end{equation}

\section{Testing the hydrodynamic theory by Monte-Carlo methods.}\label{MCtesting}

Equations~(\ref{obdm-final-meso})-(\ref{n0n-final-macro}) constitute the main result of our work and allow us to calculate $g_1({\bf r})$ (a non-diagonal function) relying only on the knowledge of the static structure factor (a diagonal function) at zero temperature $S({\bf p})$. In doing so, we assume that the superfluid fraction is large and neglect the normal component. In this way, our approach is applicable even if the condensate fraction is tiny thus providing predictions in parameter regions characterized by the failure of Gross-Pitaevskii, Bogoliubov, and other theories which are perturbative in the non-condensed occupation. In order to verify the correctness of our theory we compare the results obtained for $g_1({\bf r})$, according to the prescription given above, and the exact results obtained in non-perturbative quantum Monte Carlo methods. We use the most appropriate QMC methods to calculate the OBDM, that is diffusion Monte Carlo (DMC) and path-integral ground state (PIGS) methods at zero temperature, and path-integral Monte Carlo (PIMC) at finite temperature. 

\begin{figure}[tb]
\centering
\includegraphics[width=\columnwidth]{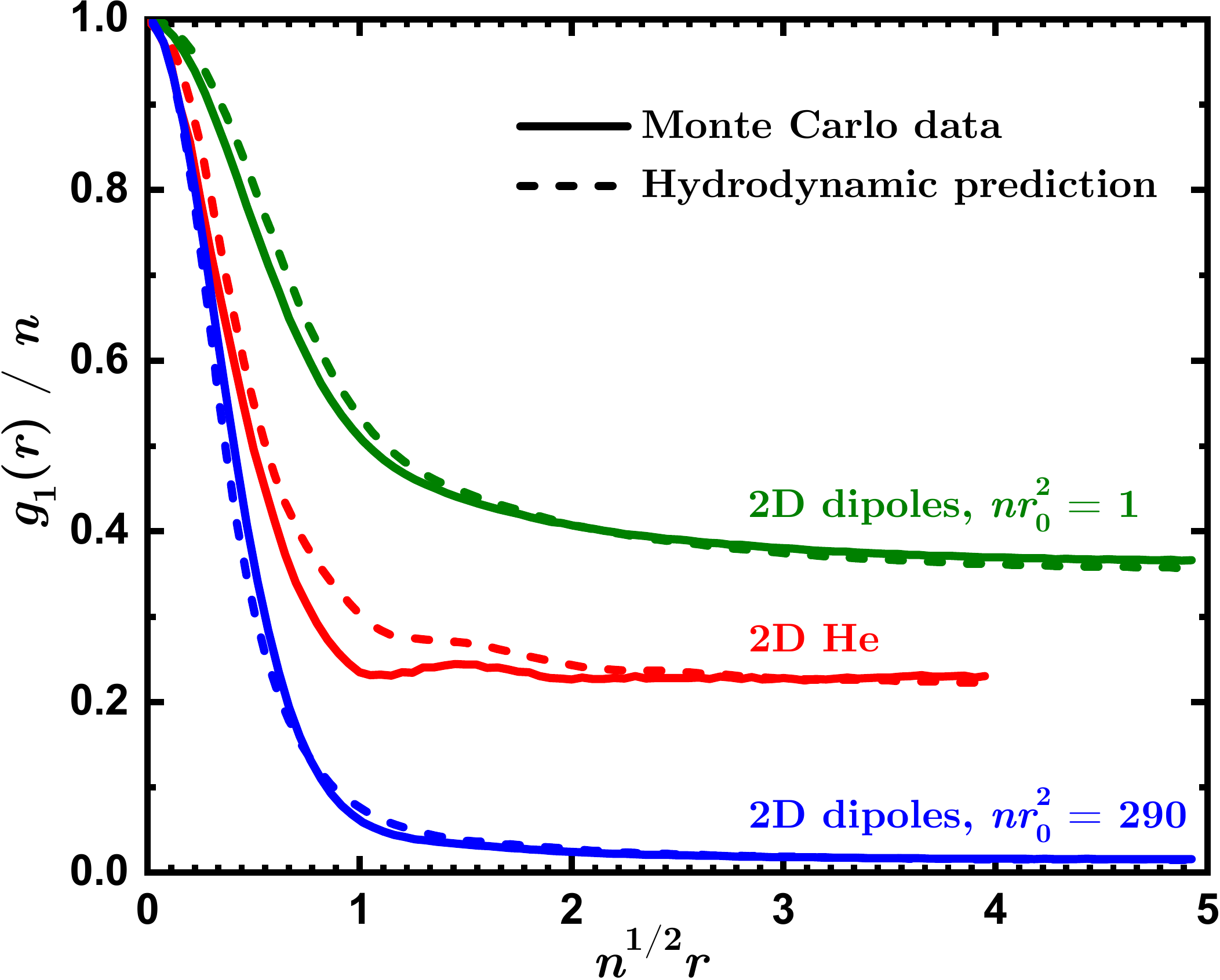}
\caption{
The one-body density matrix $g_1({\bf r})$ in 2D strongly-correlated quantum systems at zero temperature.  The solid lines are QMC results. Upper line: DMC results obtained for dipoles at density $nr_0^2=1$ and $N=100$ particles; middle line: PIGS results for $^4$He at $n=0.04347$ \AA$^{-2}$ and $N=64$; lower line: variational Monte Carlo (VMC) results for dipoles at $nr_0^2=290$ and $N=100$. The dashed lines are the  predictions of HD theory.
} 
\label{fig1}
\end{figure}

In Fig.~\ref{fig1}, we show a comparison of the OBDM for the ground state of some 2D systems. The path-integral ground state (PIGS) method is used for 2D liquid $^4$He at its equilibrium density and the DMC method for 2D quantum dipoles, with all the dipolar moments oriented perpendicularly to the plane, at densities $nr_0^2=1$ and $nr_0^2=290$ (dipolar length $r_0$ is defined as in Ref.~\cite{prl098060405}). As it is known, in the $T=0$ limit a 2D quantum fluid has a finite condensate fraction or, in other words, it manifests off-diagonal long-range order. In the same figure, we report the results obtained for the same conditions and system using HD theory. We emphasize that our analytic approach requires only the knowledge of the static structure factor at zero temperature. To this end, we use the function $S({\bf p})$ provided by the same QMC methods used to estimate $g_1({\bf r})$. As anticipated, our approach is accurate for small ${\bf p}$, or equivalently for large ${\bf r}$. And this is, in fact, observed in Fig.~\ref{fig1} where predictions of hydrodynamics are compared with the exact QMC results. Notably, we find out that even at intermediate and smaller distances both results are not so different, especially for the dipolar system.

Focusing on the limit of large distances, we can estimate the condensate fraction from the long-range asymptotic limit of $g_1({\bf r})$ as $n_0\approx g_1(L/2)$ where $L=(N/n)^{1/D}$ is a box size. Our HD theory yields results for $n_0/n$ which match the QMC ones within their statistical error. We obtain $n_0/n=0.22$ for $^4$He and $n_0/n=0.36$ and $0.015$ for dipoles at densities $nr_0^2=1$ and 290, respectively. Remarkably, even for such small values of the condensate fraction, corresponding to strongly interacting systems, we obtain a perfect agreement. It is worth mentioning that standard perturbative theories are not applicable for such strongly-correlated systems.

\begin{figure}[tb]
\centering
\includegraphics[width=\columnwidth]{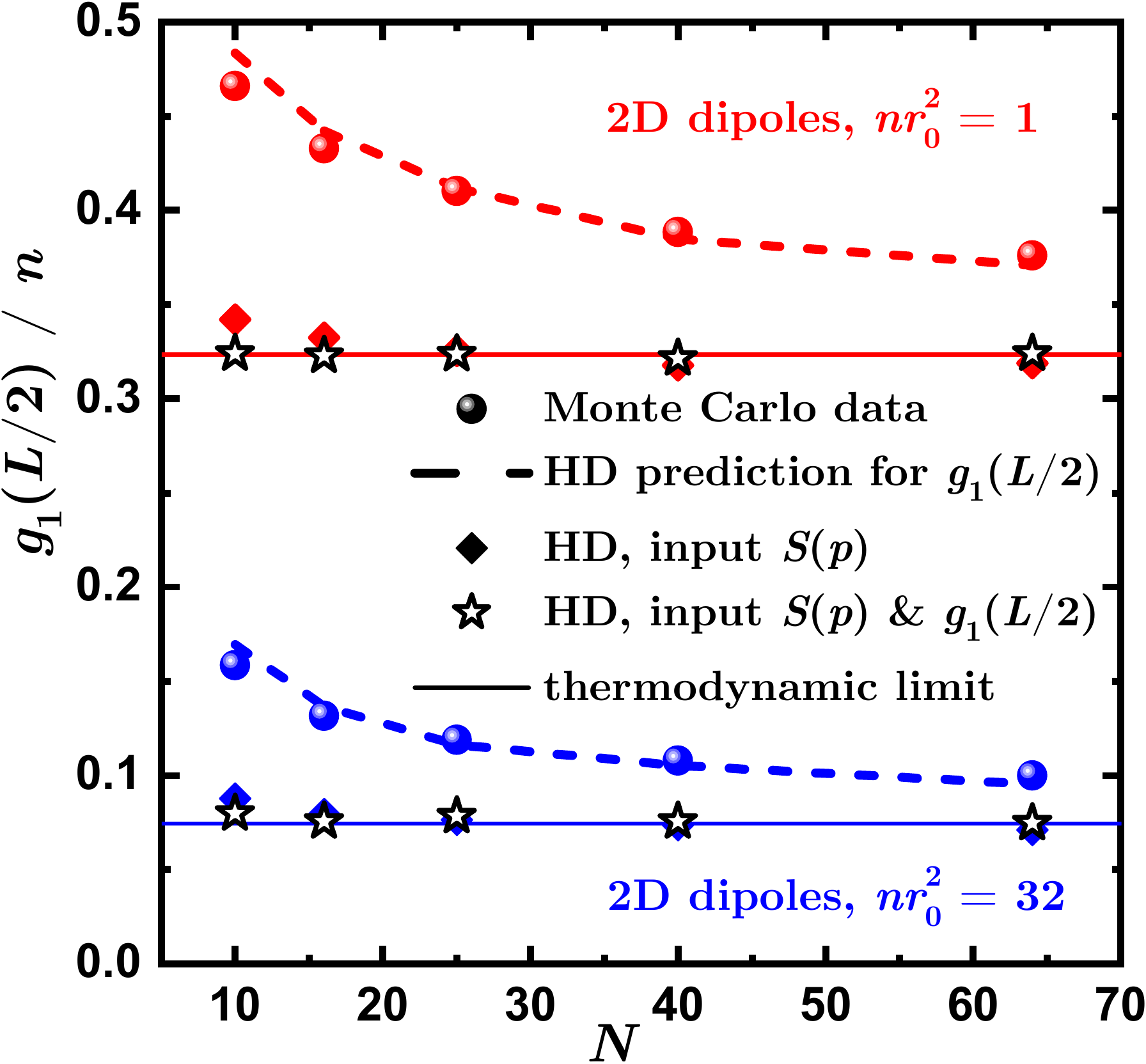}
\caption{The one-body density matrix at $r=L/2$, as a function of the particle number $N$ and at two different interaction strengths. The circles stand for the DMC results and the dashed lines correspond to finite-size HD theory. Diamonds show the results of extrapolation from the finite-size data to the thermodynamic limit using the HD theory with $S({\bf p})$ as input data, Eq.~(\ref{n0n-final-macro}). Stars use $S({\bf p})$ and $g_1(L/2)$ as input, Eq.~(\ref{SM:ansatz}). The solid lines are the condensate fractions. The data correspond to $T=0$ 2D dipoles at densities $n r_0^2=1$ (top) and 32 (bottom).}
\label{fig1c}
\end{figure}

It can be seen from Fig.~\ref{fig1} that the OBDM is correctly described by HD theory already at small distances, $r\gtrsim \ell n^{-1/D}$, starting from $\ell \approx 1.5 $--$3$ interparticle separations. Hence, the HD approach is expected to be valid in mesoscopic systems, i.e., for numbers of particles starting from $N\sim\pi\ell^2 \approx 10$--$30$ in a 2D geometry and $N\sim 4\pi\ell^3/3 \approx 30$--$100$ in a 3D one. This hypothesis is verified in Fig.~\ref{fig1c}, which shows the comparison of HD theory predictions with the results of first-principle simulations. Furthermore, the small offset remaining at large distances can be significantly reduced if the exact value of the OBDM is known in a single point for a sufficiently large distance $r^*$ at a low temperature for a large number of particles. In this case, one can use the ansatz $g_1(r) = [g_1^{\rm HD}(r)]^{1+\varkappa}$ to describe all other distances, temperatures, and particle numbers, with the exponent $|\varkappa|\lesssim 0.03$ fixed by matching the HD prediction~(\ref{obdm-final-meso}) $[g_1^{\rm HD}(r^*)]^{1+\varkappa}$ to the exact value $g_1(r^*)$. Notably, the thermodynamic value of the condensate fraction is rather accurate even when it is obtained from $S({\bf p})$ and $g_1(r^*)$ calculated in systems containing as few as 10 particles. Moreover, starting from 30--40 particles, the prediction for its thermodynamic value becomes almost exact. Thus, we conclude that our HD theory captures correctly the behavior of the OBDM already for small systems sizes (see Fig.~\ref{fig1c}) at which the OBDM is not yet converged to its thermodynamic shape. The success of the HD theory in predicting the OBDM in small systems can be understood by the fact that the static structure factor typically converges much faster to the thermodynamic limit than the OBDM itself. For example,  it can be shown that, within the HD approach in two dimensions, $g_1(L/2)$ converges to $n_0$ as $O(1/L)$, i.e., $O(1/\sqrt N)$ in terms of the number of particles. 

\begin{figure}[tb]
\centering
\includegraphics[width=\columnwidth]{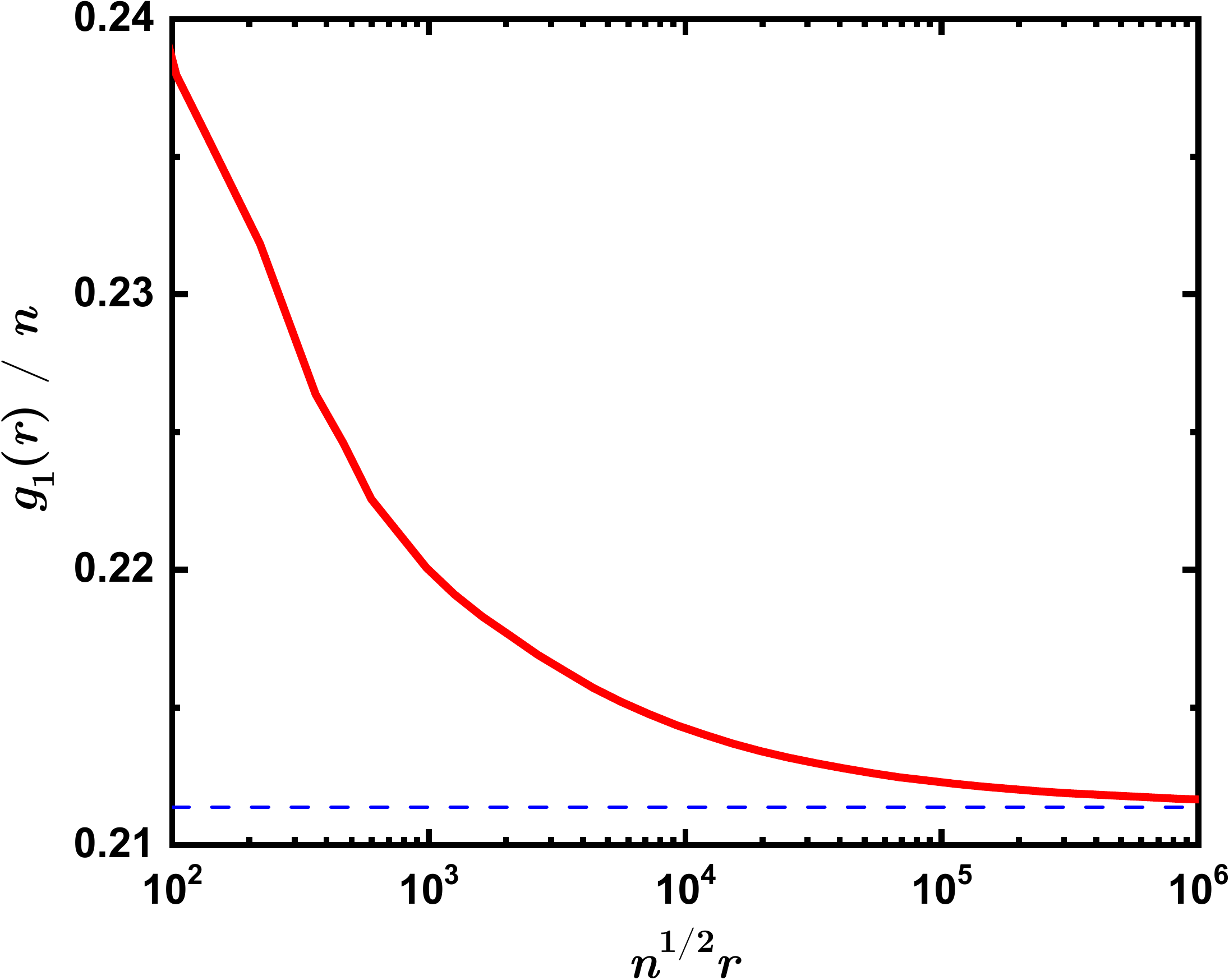}
\caption{HD prediction for the OBDM at $r=L/2$ as a function of the particle number $N$ in 2D $^4$He at the equilibrium density, $n=0.04347$ \AA$^{-2}$, and $T=0$ (red line). It is compared to its asymptotic value (dashed line), Eq.~(\ref{n0n-final-macro}). All curves are based on the PIMC static structure factor calculated at 500~mK and $N=64$, and considering a linear phononic behavior for small momenta.
}
\label{fig3main}
\end{figure}

We envision a practical utility of our method in predicting the OBDM for large system sizes, which are hard or even impossible to simulate directly. As an example, we can calculate the OBDM in 2D $^4$He at zero temperature for system sizes as large as one million particles. Fig.~\ref{fig3main} shows the convergence of the condensate fraction to the macroscopic limit in this case. Even for systems consisting of thousands of particles, the difference between the finite-size condensate fraction and its thermodynamic value is significant. Only for system sizes as large as $N\approx 10^6$ the differences become negligible. It is appropriate to note that a direct simulation with any QMC method consisting of $10^6$ particles, if possible, would require the use of large supercomputing facilities. On the contrary, the use of our extrapolation method yields a high-precision estimation at the thermodynamic limit already starting from $N\approx100$, as shown in Fig.~\ref{fig1c}.

\begin{figure}[tb]
\centering
\includegraphics[width=\columnwidth]{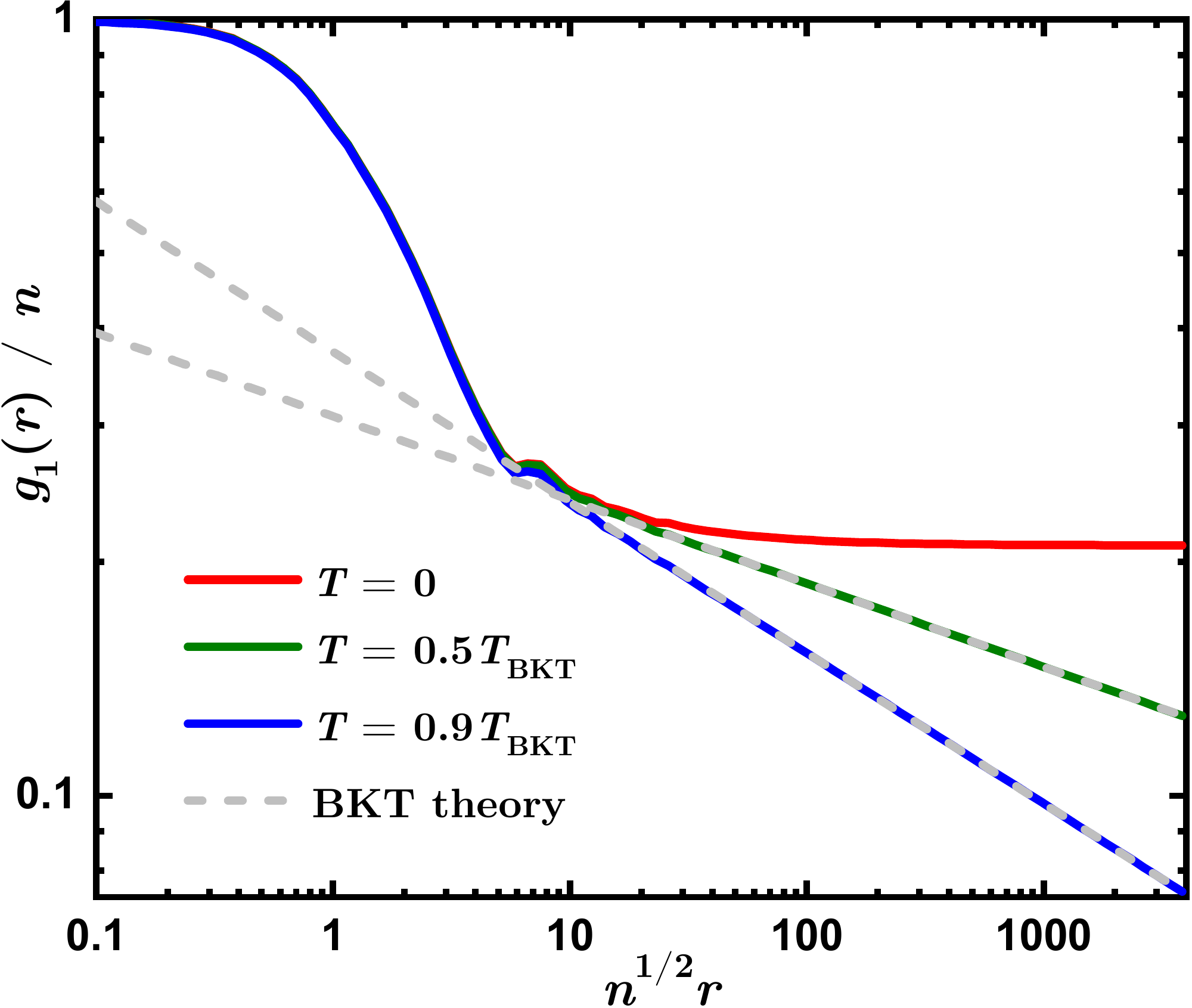}
\caption{The one-body density matrix $g_1({\bf r})$ for 2D liquid $^4$He and same parameters as in Fig.~\ref{fig3main} as obtained from the HD method, at different temperatures, expressed in units of the BKT critical temperature $T_{\text{BKT}}$. Dashed lines show the long-range power-law asymptotic behavior predicted by the BKT theory below $T_{\text{BKT}}$.} \label{fig2}
\end{figure}  

Interestingly, our approach is able to describe correctly the effect of a finite temperature on spatial coherence. The advantage of using HD theory is that it properly takes into account both thermal and zero-point phase fluctuations. In Fig.~\ref{fig2}, we show the long-range behavior of $g_1({\bf r})$ for 2D $^4$He at temperatures below the BKT critical temperature. It is worth noticing that, to produce these results, we only need the zero-temperature static structure factor calculated for a single finite size of the simulated system, with periodic boundary conditions. As one can see, our results perfectly match the expected BKT power-law decay, $g_1({\bf r}) \sim r^{-\alpha}$, with $\alpha= mT/(2 \pi \hbar n_s)$.

\begin{figure}[tb]
\centering
\includegraphics[width=\columnwidth]{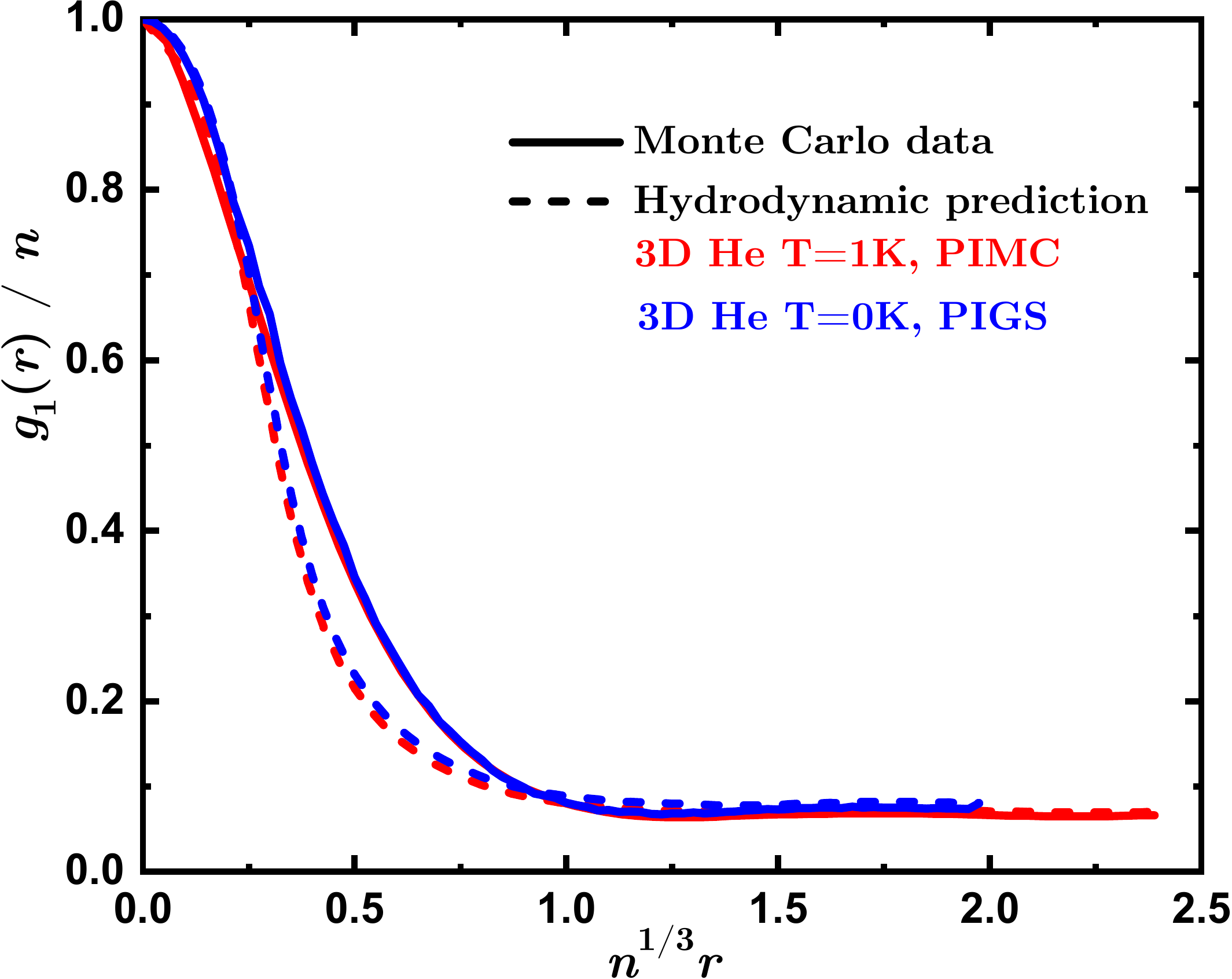}
\caption{The one-body density matrix $g_1({\bf r})$ in bulk $^4$He at zero and finite temperatures. Solid and dashed lines stand for QMC and HD results, respectively.} 
\label{fig3}
\end{figure}

We verify that our method works also correctly in 3D quantum liquids. For that purpose, we make a comparison for bulk superfluid $^4$He, where we know that the condensate fraction is small due its strongly interacting nature. We show a comparison between QMC and HD in Fig.~\ref{fig3}. The results corresponding to $T=0$ are obtained with PIGS and the ones at $T=1$ K with PIMC. The equilibrium density, $n=0.02186$ \AA$^{-3}$, is used in both cases. As one can see, the agreement in the plateau is excellent and quite good at short and intermediate distances. The condensate fraction obtained with the two methods is again in agreement within the error bars of the QMC estimation, $n_0/n \simeq 0.07$. 

\begin{figure}[tb]
\centering
\includegraphics[width=\columnwidth]{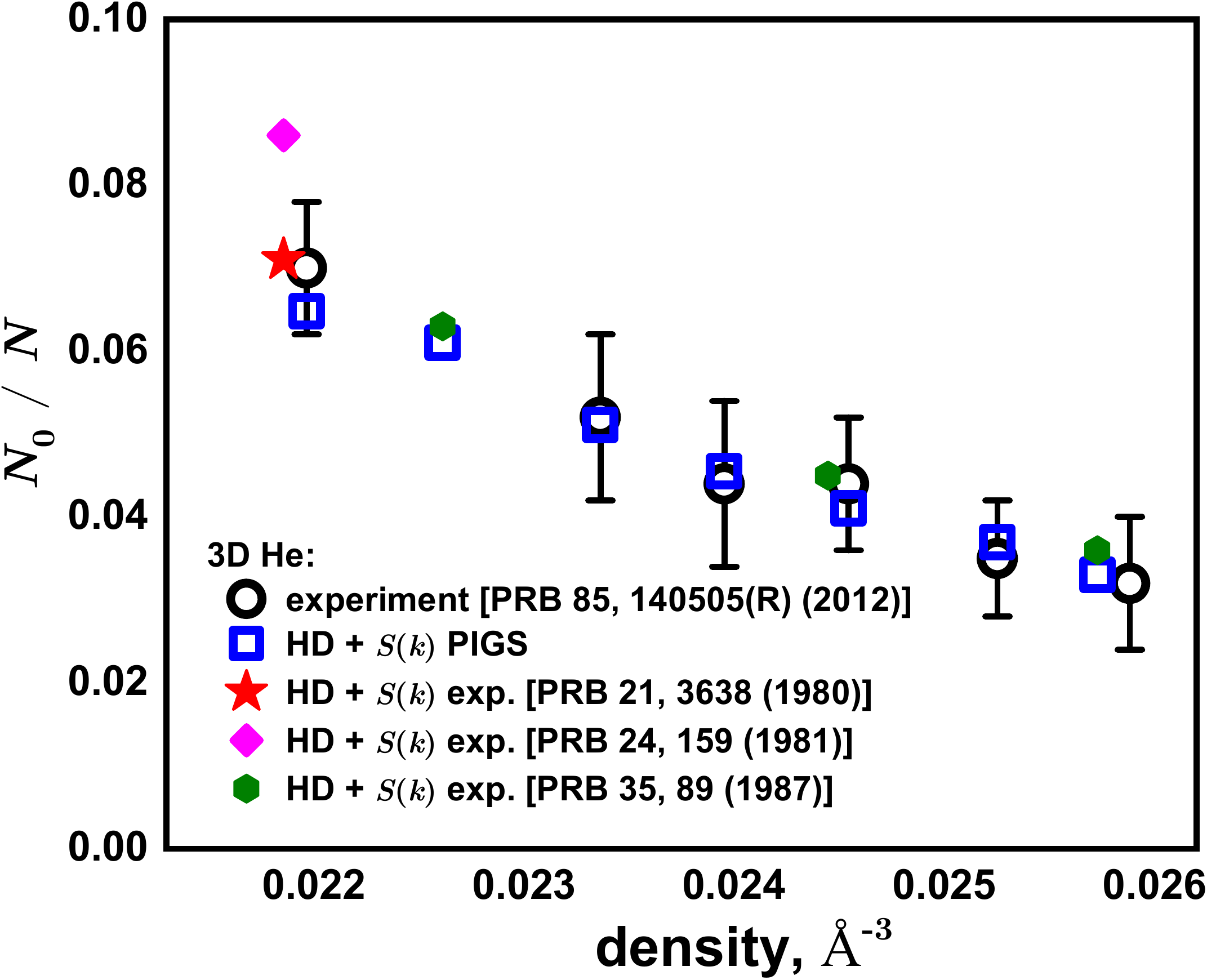}
\caption{Condensate fraction of superfluid $^4$He as a function of the density. Open circles are experimental results obtained from deep inelastic neutron scattering~\cite{glyden0}. Squares are HD predictions at the same densities. The star, diamond, and hexagonal points correspond to HD results using the experimental measurements of $S({\bf p})$ from Refs.~\cite{svensson}, \cite{prb024000159}, and \cite{wirth}, respectively.} 
\label{fig4}
\end{figure}

The comparison between the hydrodynamic predictions for the condensate fraction and available experimental results~\cite{glyden0} is the more exigent test to conclude on the usefulness of our method. This check can be made for the paradigmatic case of liquid $^4$He. In Fig.~\ref{fig4}, we report results of $n_0/n$ for superfluid $^4$He as a function of the density. The functions $S({\bf p})$, required for the one-body density matrix calculation, are obtained using the PIGS method. As we can see in the figure, an overall agreement is achieved with the experimental results~\cite{glyden0}, obtained by using deep-inelastic neutron scattering, and after an accurate inclusion of FSE in the analysis. Remarkably, the accuracy of our theory does not fall even when the freezing density is approached and the condensate fraction is significantly reduced. 

One may wonder whether the HD approach can also provide an alternative way of measuring the condensate fraction instead of the present way, based on deep inelastic scattering at large momentum transfer. In other words, whether the experimental measure of the static structure factor can be used to infer $n_0/n$. We have verified this possibility by using in our theory three independent experimental measurements of $S({\bf p})$, two obtained by x-rays~\cite{wirth,prb024000159} and the other by neutron scattering~\cite{svensson} at low momentum transfer. Our method predicts $n_0/n=0.071$ based on the experimental $S({\bf p})$ from Ref.~\cite{svensson}, and $n_0/n=0.086$ for data from Ref.~\cite{prb024000159} (see Fig.~\ref{fig4}). Unfortunately, Ref.~\cite{prb024000159} does not report the values of $S({\bf p})$ for large momenta, resulting in an overestimation of $n_0/n$ due to the use of truncated data in the HD analysis. Data form Ref.~\cite{wirth} produces results at different densities that are very close to the results from Ref.~\cite{glyden0}, obtained through deep inelastic neutron scattering, thus validating the usefulness of our method for extracting the condensate fraction in an alternative way.

\section{Criteria for a proper cut-off function}\label{cutoff}

The proper choice of the cut-off function $\kappa_{\bf p}({\bf r})$ is central to our theory. Although we do not provide a rigorous derivation, the following considerations apply:
\begin{itemize}
\item[(a)] In the regime of weak correlations a proper choice is $\kappa_{\bf 
p}({\bf r}) = |1-S({\bf p})|^2$, with the quadratic dependence coming from the 
Bogoliubov theory~\cite{PhysRevLett.121.235702}.
\item[(b)] In the regime of strong correlations one has $\kappa_{\bf p}({\bf r}) 
=|1-S({\bf p})|$, with the linear dependence demonstrated in 
Ref.~\cite{Lozovik2008}.
\item[(c)] More generally, we assume that the cut off factor has a power-law 
form $|1-S({\bf p})|^\nu$, where the exponent must be equal to $\nu=1$ in the 
limit of strong correlations ($n_0/n=+0$) and $\nu=2$ in the limit of weak 
interactions ($n_0/n=1$). 
\item[(d)] In addition, the exponent $\nu$ is assumed to be a function of the condensate fraction, $\nu = \nu(n_0/n)$.
\item[(e)] We demonstrate that an excellent agreement with experimental data and 
results of numerical simulations is obtained when the cut-off function is chosen 
as $\nu(x)=2/(2-x)$. The high quality of the predictions was illustrated in 
Sec.~\ref{MCtesting} and a comparison with a number of different cutoffs is 
shown in Sec.~\ref{optimal-cutoff}.
\item[(f)] Once the cut-off expression has been chosen, we have verified that the OBDM and mesoscopic values of the condensate fraction are correctly reproduced. Here, the underlying assumption is that instead of the thermodynamic condensate fraction $\nu(n_0/n)$ one should use the value of the OBDM at half size of the box $\nu(g_1(L/2)/n)$.
\item[(g)] Finally, we use scaling considerations (similar to those used in BKT theory\cite{Kosterlitz73}) and we substitute $L/2$ by $r$ in the OBDM. This leads to the final form of the cut off as given in Eq.~(\ref{SM:kpr}). In other words, for a given length $r$ we do the following: (i) instead of the original system we consider a different one of size $L\sim r$ and (ii) we approximate $g_1(r)/n$ by the condensate fraction in the system of size $L\sim r$.
The obtained expression does not apply to short distances as $g_1$ has a fast dependence on $r$ and hydrodynamics does not apply at distances smaller than the mean interparticle distance. 
A similar method was used\cite{Lozovik2007} to calculate the superfluid fraction $n_s / n$ in a mesoscopic system of a finite size $L$ from the known scale-dependent value of the superfluid fraction $n_s(r) / n$ in the thermodynamic limit.
\item[(h)] In two dimensions, there is an infrared Hohenberg divergence\cite{pr0158000383} at $T\ne 0$. Therefore, we use $g_1^{T=0}({\bf r})$ rather than $g_1({\bf r})$ in the exponent~(\ref{SM:kpr}). Indeed, because of the power-law decay in the one-body density matrix \cite{pr0155000080} $g_1({\bf r})\propto r^{-mT/(2\pi\hbar^2n_s)} \to 0$ for $r\to\infty$ we would obtain $2/(2-g_1({\bf r})/n)\to1$ for all cases. Whereas according to us, the correct value of the exponent in Eq.~(\ref{SM:kpr}) at $r\to\infty$ must be different from unity and equal to $2/(2-n_0^0/n)$, where $n_0^0=g_1^{T=0}({\bf r}\to\infty)$ is the quasi-condensate density, i.e., the condensate density at $T=0$.
\end{itemize}

Based on the above considerations, we chose the cut-off factor in the form given in Eq.~(\ref{SM:kpr}). 
The ultimate test for the accuracy of the proposed theory is the verification of the method in a direct comparison with the experimental and theoretical results for the static structure factor $S({\bf p})$ and the OBDM $g_1(r)$.

\section{Optimal cut-off function}\label{optimal-cutoff}

It is important to verify the stability of our method with respect to a particular choice of the cut-off function. To do so we have considered a number of cut-off functions, $\nu(x)$ with $x = n_0/n$, satisfying the conditions discussed in Sec.~\ref{cutoff}. In particular, the following eight choices have been used: 
\begin{equation}
\begin{array}{llll}
\mbox{1) }&\nu(x)&=&2/(2-x^2),\\
\mbox{2) }&\nu(x)&=&1+x^2,\\
\mbox{3) }&\nu(x)&=&2/(2-x)\mbox{ optimal},\\
\mbox{4) }&\nu(x)&=&2^x,\\
\mbox{5) }&\nu(x)&=&1+x,\\
\mbox{6) }&\nu(x)&=&1/(1-x+x^2/2),\\
\mbox{7) }&\nu(x)&=&4^{x^2/(1+x^2)},\\
\mbox{8) }&\nu(x)&=&1+x(2-x).
\end{array}
\label{Eq:cutoffs}
\end{equation}
In particular, all considered choices of the cut-off function satisfy the conditions for weak correlations, $\nu(1)=2$, and strong correlations, $\nu(0)=1$.

\begin{figure}[tbh]
\centering
\includegraphics[width=\columnwidth]{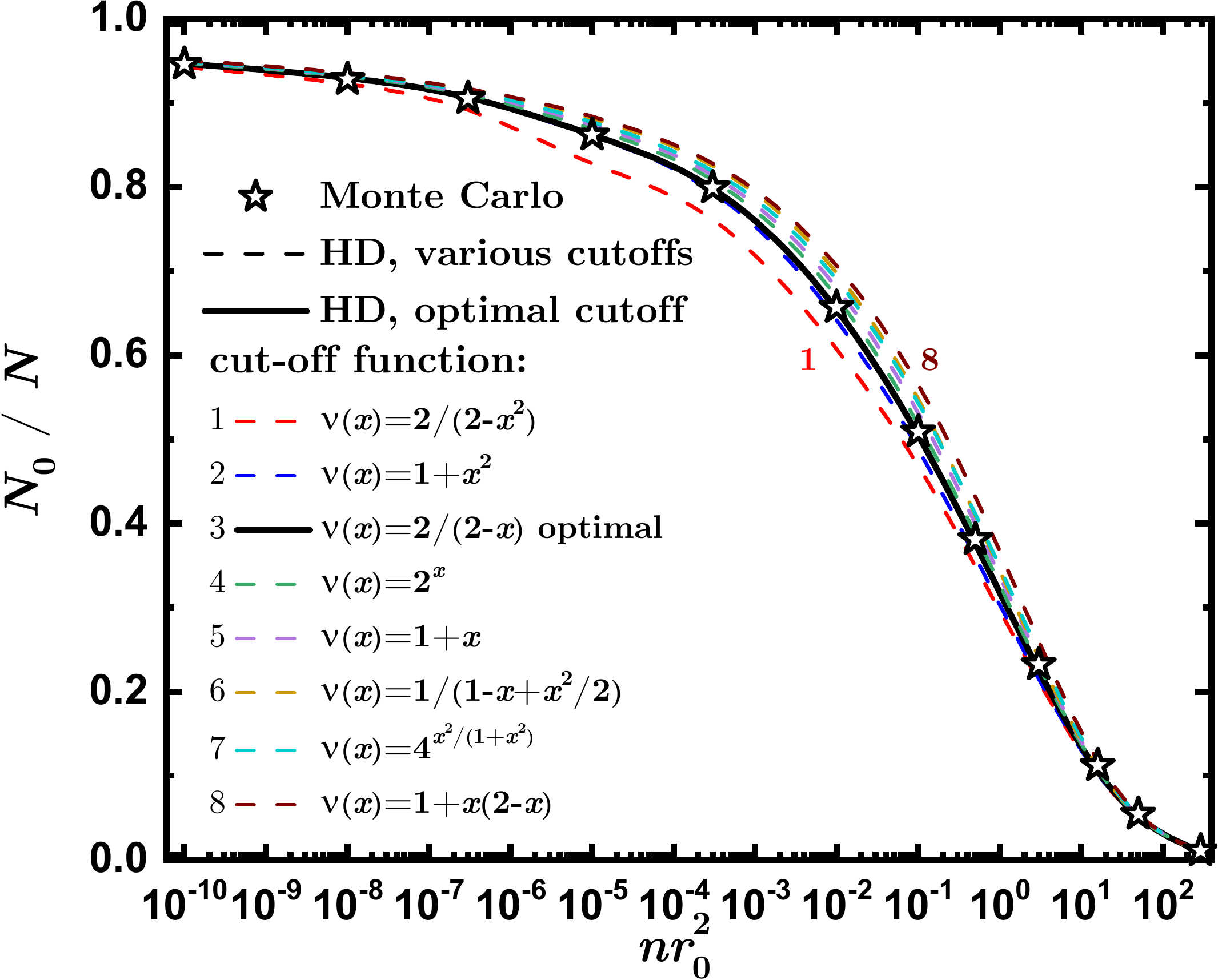}
\caption{
Dependence of the accuracy of the HD prediction for the macroscopic condensate fraction $N_0/N$ on the choice of the cut-off function $\nu(x)$ for 2D dipoles as a function of the density $nr_0^2$.  Stars, condensate fraction obtained directly and used as a reference (QMC+HD theory based on OBDM and $S({\bf p})$ calculated with $N=100$ particles, VMC+DMC extrapolation is done for $N_0/N$ values); dashed lines, predictions of our hydrodynamic method with 8~different cut-off functions (Eq.~(\ref{Eq:cutoffs}) in increasing order from bottom to top); solid line, prediction of our hydrodynamic method with the optimal cut-off function $\nu(x) = 2/(2-x)$.
}\label{figCutOff}
\end{figure}

In order to verify the accuracy of different cut-off functions, Eq.~(\ref{Eq:cutoffs}), we use a 2D dipolar system as a reference. Figure~\ref{figCutOff} shows the comparison of the condensate fraction as we get it from the hydrodynamic theory with the reference data, obtained by extrapolating the OBDM to the thermodynamic limit. We consider a wide range of densities, $10^{-10}<nr_0^2<290$, covering a variety of regimes from weakly-interacting Bogoliubov gas and up to the highest possible density after which the dipolar gas experiences a transition to a solid.

It can be seen that there is a negligible dependence on a specific choice of $\nu(x)$ both in the Bogoliubov perturbative regime, $N_0/N\gtrsim 90\%$, and in the regime where the condensate fraction is almost completely exhausted, $N_0/N\lesssim 10\%$. Those are the regimes where all considered functions approach appropriate the limiting values which are known analytically. This is not the case for the intermediate strength of correlations, where there is a weak dependence on the specific choice of $\nu(x)$. Indeed, the maximum difference between different choices observed for $N_0/N\approx 60\%$ is smaller than $10\%$. 

Out of the considered functions, the choice $\nu(x)=2/(2-x)$ stands out for the 
high quality of the results obtained and so it is the one used in our theory.

\section{Improved extrapolation procedure}\label{extra-ansatz}

As shown above, the hydrodynamic theory can be used to predict the long-range behavior of the OBDM based on the knowledge of $S({\bf p})$. At the same time, the hydrodynamic theory fails at short distances and hence it introduces a small offset in the OBDM at large distances. The value of the offset depends on details of the pair interactions, density, dimensionality, etc. This offset can be removed if a single value of the OBDM is known. That is if at least a single value of $g_1(r^*)$ is known at some large distance $r^*$, then it is possible to predict well the long-range behavior of OBDMs for any low temperature, large distance and large particle number. In particular, such a procedure is especially important for extrapolation of the condensate fraction to the thermodynamic limit in numerical simulations.

An improved procedure relies on the simultaneous knowledge of $S({\bf p})$ for different values of momentum $p$ and $g_1(r)$ for a single point $r=r^*$ (e.g. $r^*=L/2$) in a finite-size system at some (zero or low) temperature. Then the finite-size effects can be significantly reduced by approximating the $g_1(r)$ by the ansatz
\begin{equation}
g_1(r) = [g_1^{\rm HD}(r)]^{1+\varkappa}
\label{SM:ansatz}
\end{equation}
where $g_1^{\rm HD}(r)$ is given by Eq.~(\ref{obdm-final-meso}) and $\varkappa$ is chosen in such a way that $g_1(r^*) = [g_1^{\rm HD}(r^*)]^{1+\varkappa}$ for the distance $r=r^*$ at which the exact value of OBDM is known. In this way, the long-range envelope of OBDM follows the form predicted by hydrodynamics while the offset is essentially removed by appropriately adjusting the value of $\varkappa \ll 1$. We note that it is convenient to use a power-law ansatz instead of an additive shift in order to preserve the correct value of the OBDM at origin as $g_1(r=0) = 1$.

We propose the following extrapolation procedure:
\begin{itemize}
\item[(1)] measure $S({\bf p})$ for different momenta and $g_1(r)$ in a single 
point $r^*$,
\item[(2)] calculate the OBDM $g_1^{\rm HD}(r)$ with HD theory using $S({\bf 
p})$ as an input,
\item[(3)] fix the value of parameter $\varkappa\ll 1$ in such a way that the 
ansatz~(\ref{SM:ansatz}) is exact for the point $r^*$ in which the OBDM is 
known,
\item[(4)] then, the ansatz~(\ref{SM:ansatz}) can be used for any sufficiently 
low temperature $T$, large number of particles $N$ and large distance $r$.
\end{itemize}
Typically, the adjustable parameter $\varkappa$ is of order of $3$\%. Its specific value depends on the interaction type, density, system dimensionality, etc.  At the same time, its value is almost independent of the distance $r^*$, particle number and temperature if $T$ is sufficiently small while $r^*$ and $N$ are large enough.

It is worth noting that the ability of predicting the finite-size results is especially important in two dimensions where the condensate fraction is absent at any finite temperature.

\section{Finite-size effects}\label{finite}

\begin{figure}[tbh]
\centering
\includegraphics[width=\columnwidth]{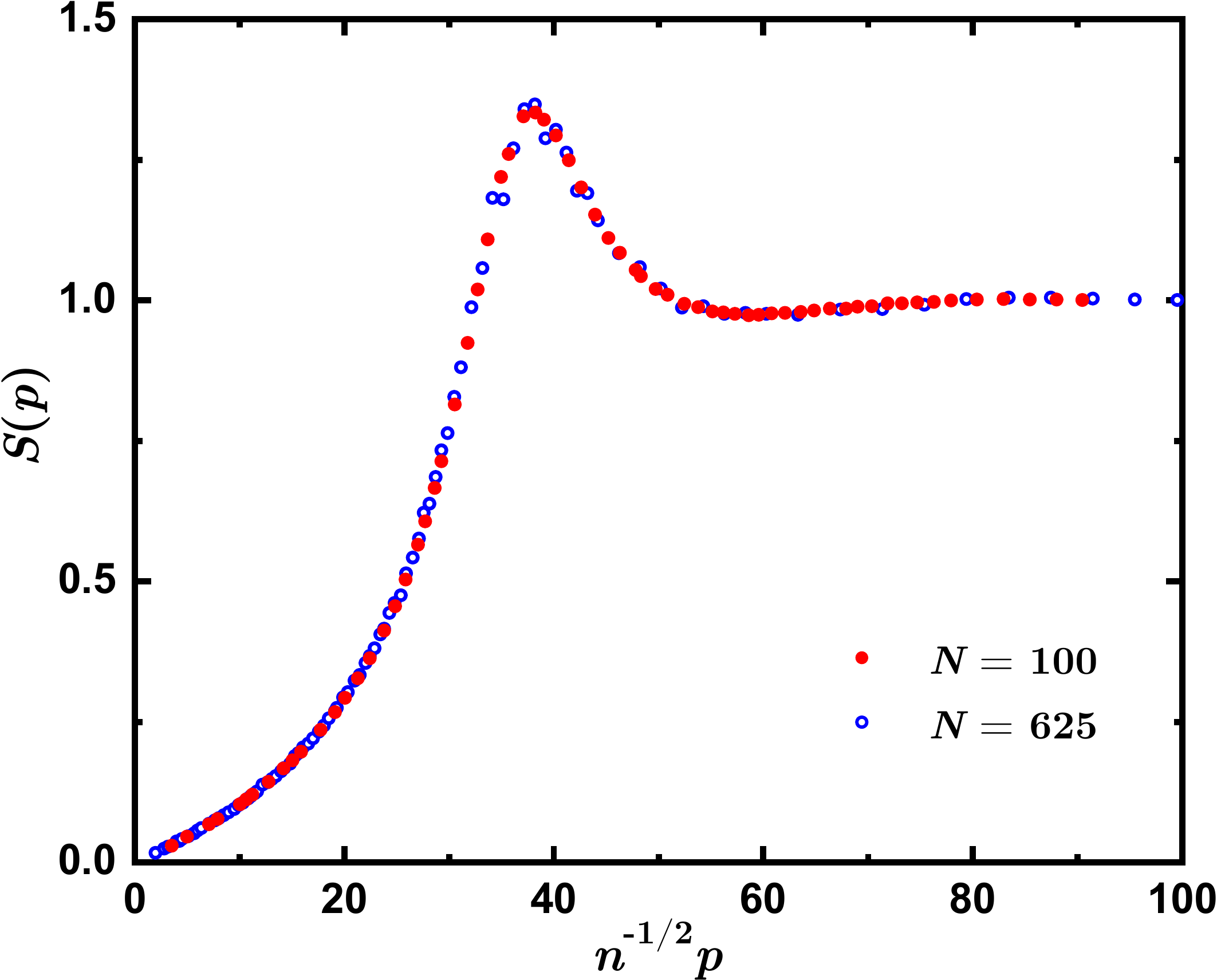}
\caption{
An example of finite-size effect in the static structure factor for $N=100$ and $N=625$ particles in a two-dimensional dipolar system at dimensionless density $nr_0^2=32$.
} 
\label{fig7}
\end{figure}

The advantage of using the static structure factor $S({\bf p})$ as an input is that typically it has much faster convergence to the thermodynamic limit as compared to the convergence of the OBDM $g_1(r)$.  Indeed, the main effect of a finite-size box on the static structure factor is the discretization of the allowed momenta. The value of the low-momentum linear slope is related to the speed of sound which is almost not changed with system size. As a result, the overall shape of the static structure factor does not significantly change with the number of particles (see Fig.~\ref{fig7} for a characteristic example). If the size of the box is larger than the mean interparticle distance, the missing small-momentum behavior can be safely reconstructed from the linear phonons, $S({\bf p}) = p/(2mc)$ where $c$ is the speed of sound. 

\begin{figure}[tbh]
\centering
\includegraphics[width=\columnwidth]{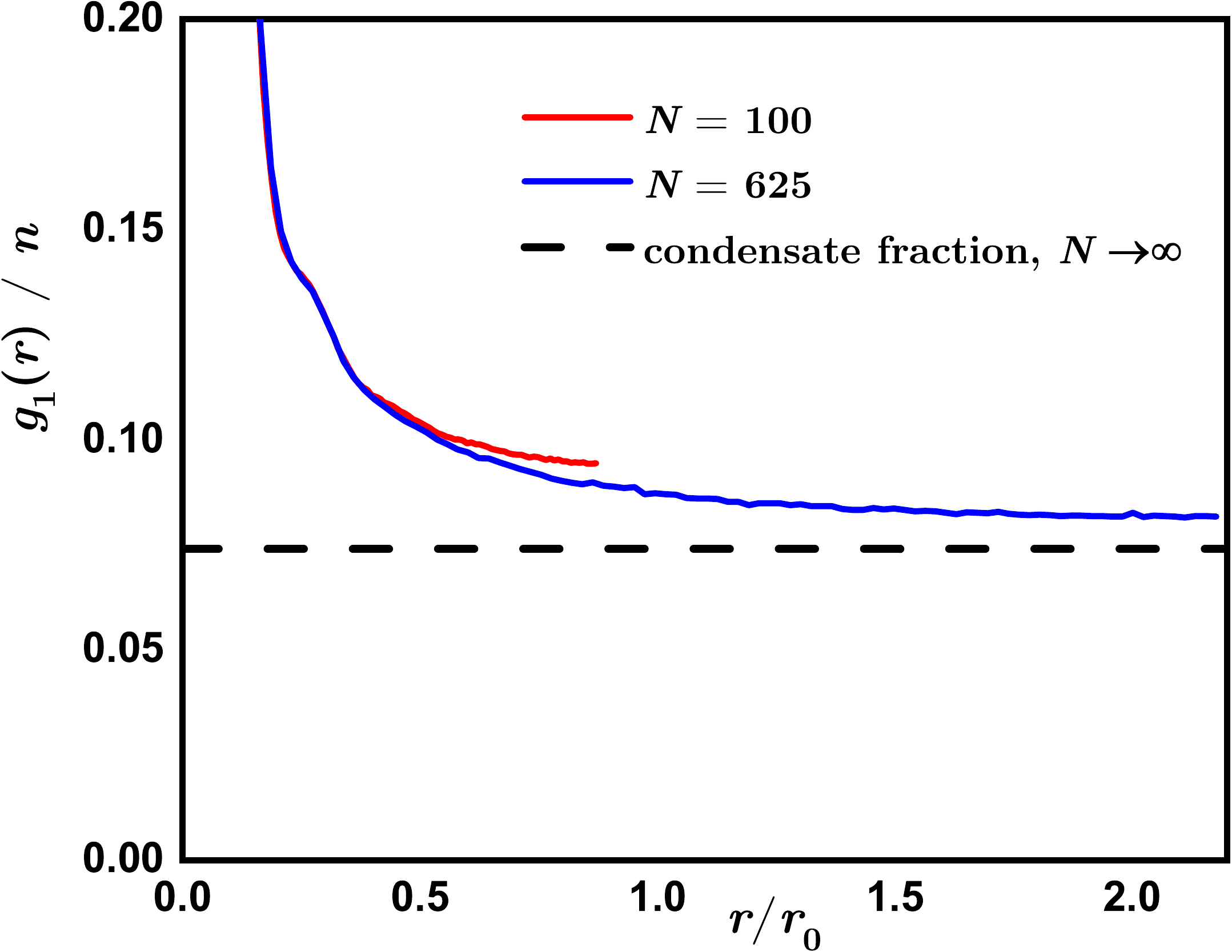}
\caption{
An example of finite-size effect in the one-body density matrix for $N=100$ and $N=625$ particles in two-dimensional dipolar system at dimensionless density $nr_0^2=32$ .
} 
\label{fig8}
\end{figure}

Instead, the change of the OBDM as the size of the box is increased can be significant, see Fig.~\ref{fig8}. There are several reasons for this much stronger dependence. 

One reason is that according to the periodic boundary condition the OBDM $g_1({\bf r})$ must be a periodic function of ${\bf r}$ when any of the $x,y,z$ arguments is displaced by the box length $L$. This flattens the OBDM at distances $r\approx L/2$, for example, in a one-dimensional geometry the first derivative $g'(L/2)=0$. 

Another reason is the exponential dependence of the OBDM on phase fluctuations, see the cumulant Kubo Eq.~(\ref{obdm1}). The stronger are the correlations, the larger are the speed of sound and the coefficient in the exponent according to Eq.~(\ref{obdm-final-meso}) for $S(p) \propto 1/v_s$. This makes convergence with $N$ even slower. In contrast, there is no cumulant exponent in the structure factor which is rather the Fourier transform of the pair correlation function (for details on cumulant technique refer to Ref.~\cite{pr0155000080}). Gapless systems at zero temperature feature a fast power-law decay in the long-range asymptotic of the pair distribution function as $1/r^{D+1}$ due to the phononic behavior of $S(p) \propto p$ at small momenta $p$. The decay in OBDM is slow and $g_1(r)$ scales as $1/r^{D-1}$ at large distance since the momentum distribution at small $p$ behaves like $n(p) \propto 1/p$.

\begin{figure}[tbh]
\centering
\includegraphics[width=\columnwidth]{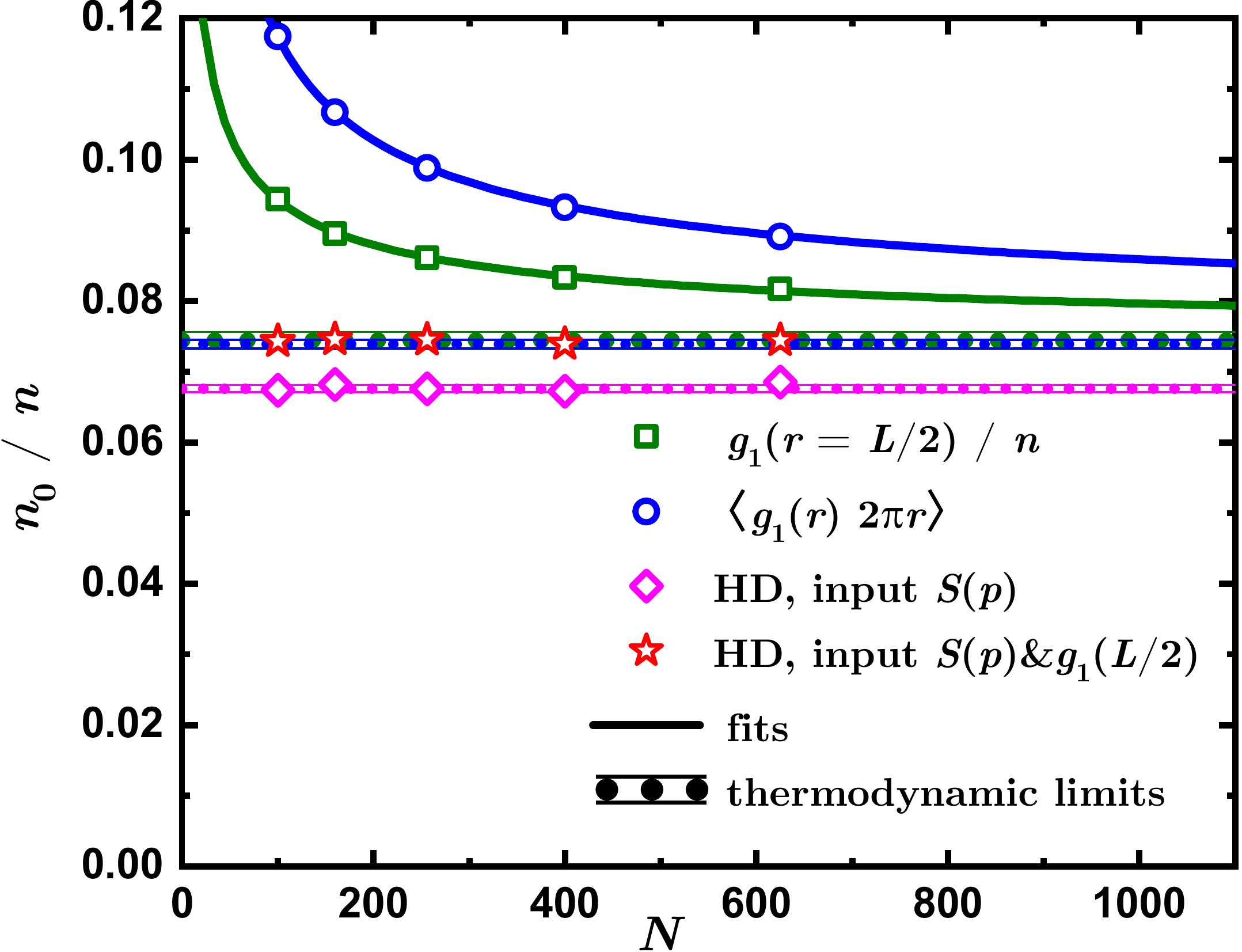}
\caption{
Example of a finite-size convergence of the condensate fraction. Squares, the one-body density matrix calculated at the largest possible distance, $g_1(r=L/2)$. Circles, average value of the OBDM $\int_0^{L/2} g_1(r)2\pi r\;dr / (\pi L^2/4)$. Diamonds, hydrodynamic theory prediction for the thermodynamic condensate fraction based on the static structure factor for a given number of particles, Eq.~(\ref{n0n-final-macro}). Stars, thermodynamic values as obtained from the ansatz~(\ref{SM:ansatz}) at $r^*=L/2$. Solid lines, fits in the form of $a+b/N^{1/2}+c/N$ where $a$, $b$ and $c$ are free parameters. Thick dashed lines, thermodynamic values as obtained from the fit.
} 
\label{figSM:CF}
\end{figure}

The mentioned properties of finite-size convergence of the static structure factor and OBDM suggest that the extrapolation to the thermodynamic limit is more efficient when it is done based on $S({\bf p})$. In Fig.~\ref{figSM:CF} we compare different methods of extrapolation to the thermodynamic limit. One method is based on the definition of the condensate number in terms of the momentum distribution $n_p = \int g_1(r) e^{i{\bf pr}}\;d{\bf r}$ as the number of particles with zero momentum $n_0 = \int g_1(r) d{\bf r}$. Thus, one integrates the OBDM and then takes the limiting value in the $N\to\infty$ limit. This method has the slowest convergence to the thermodynamic limit from the considered ones. Another method is based on the calculation of the OBDM value at the half size of the box which is the largest allowed distance, $g_1(L/2)$, and taking the thermodynamic limit. The use of our method is advantageous as (i) it shows the fastest convergence (ii) commonly the calculation for large system sizes is much more numerically expensive ($N^2$, $N^3$ or even exponential scaling is common) and even might require the use of supercomputing facilities so dealing with small system sizes is preferable (iii) there is no need to calculate a grid in the system size number.

\section{Sensitivity to the accuracy of the input data}\label{noised-data}

The convergence and accuracy of our method are sensitive to the quality of the static structure factor. Here we discuss two typical issues that might be present in $S({\bf p})$ at (i) low (ii) high momenta.

The low-momentum part of the static structure factor is defined by linear phonons. Unfortunately, it is quite common that the signal from low-energy Bragg scattering can be hardly measured in experiments. Also, numerical calculation of small momenta requires resource-intensive simulations of large system sizes at low temperatures. In addition to the need for simulation with large numbers of particles, the zero-temperature algorithms which are based on projection techniques (diffusion Monte Carlo, path-integral ground-state methods, etc) have the worst convergence for the low-energy modes. That is, even if the energy is converged, the low-momentum part of the $S({\bf p})$ still might require an additional effort to be correctly simulated. There are different possible ways of dealing with this problem. One method is to impose a proper linear dependence $S({\bf p}) = p/(2mc)$ where $c$ is the speed of sound (which can be directly measured or related to the compressibility). Another method is to remove the lowest momentum ``troublesome'' points and verify the stability of the output. Once the remaining points lie on the proper linear phononic part, the prediction of our method for the condensate fraction becomes stable. For example, using the experimental data of the Helium static structure factor from Ref.~\cite{wirth} the condensate fraction decreases from 6.6\% to 6.2\% when the first $1, 2, \cdots, 6$ points are eliminated. Instead, when the next one or several points are removed, the fraction of condensate essentially does not change, remaining equal to 6.2\%, that is the structure factor has reached the proper phonon behavior. The main differences in static structure factor between zero temperature and a finite but low one, affect the momenta with $p c \lesssim T$. Thus for practical purposes, the zero-temperature $S({\bf p})$ can be approximately obtained from $S({\bf p})$ by substitution of the low-momentum part by the proper linear phononic behavior. 

At large momenta, the static structure factor converges to a unit value. Any noise in $S({\bf p})$ in this asymptotic regime leads to an artificial lowering of the condensate fraction. The reason for that is that Eq.~(\ref{SM:kpr}) depends on $|1-S({\bf p})|$ thus fluctuations around 1 contribute to the result. A possible way out is to use a high-frequency filter or perform any other type of smoothing and to impose $S({\bf p})=1$ for all points where the difference from the asymptotic value is smaller than the error bars. For example, for the experimental data of Ref.~\cite{wirth} for the lowest temperature, the noisy structure factor gives the fraction of condensate $n_0/n=0.054$ that is very different from its fraction $n_0/n=0.062$ after the noise is removed.

\section{Conclusions}
In conclusion, we have developed a new quantum hydrodynamic theory for superfluid systems. Using our phenomenological approach for the interaction potential and the ultraviolet cutoff factor, we arrive to both finite-size and macroscopic expressions for the one-body density matrix $g_1({\bf r})$ that rely only on previous knowledge of the static structure factor at zero temperature. In this way, we provide access to a non-diagonal property starting from a diagonal one. In contrast to standard perturbation theories, which rely on a large condensate fraction and fail for a small one, the hydrodynamic method used requires a large superfluid fraction (collisionless regime) and it can be applied for large condensate depletion as well. We have verified that the values of the condensate fraction, derived from the long-range behavior of $g_1({\bf r})$, match closely the ones obtained from ab initio QMC simulations. Our approach provides an enhanced convergence of the condensate fraction to its thermodynamic-limit value because it is based on the static structure factor, which typically has reduced finite-size effects. Using the static structure factor $S({\bf p})$, obtained in a QMC calculation, we reproduce experimental results for the condensate fraction in superfluid $^4$He for a wide range of densities. Moreover, using experimental measurements of $S({\bf p})$ we obtain predictions for the condensate fraction that are statistically indistinguishable from the ones obtained by deep inelastic neutron scattering at large momentum transfer. Our method applies even to cases in which the two-body interaction potential is not exactly known (for example, for dense excitons or Rydberg gas), thus impeding a direct simulation of the system, whereas $S({\bf p})$ is experimentally accessible. It is worth noticing that recently there have been experimental measures of $S({\bf p})$ to characterize the supersolid phase in dilute dipolar systems, which can be used within our formalism to extract easily the one-body density matrix of this intriguing phase~\cite{Pfau2020,Chomaz2020}. The present theory is addressed to both experimentalists and theoreticians and it provides an alternative procedure to the usual and difficult estimation of the condensate fraction in strongly correlated quantum superfluids.

\acknowledgments
This work has been supported by the Ministerio de Economia, Industria y Competitividad (MINECO, Spain) under grant No. FIS2017-84114-C2-1-P. We acknowledge financial support from Secretaria d'Universitats i Recerca del Departament d'Empresa i Coneixement de la Generalitat de Catalunya, co-funded by the European Union Regional Development Fund within the ERDF Operational Program of Catalunya (project QuantumCat, ref. 001-P-001644). The authors thankfully acknowledge the computer resources at Cibeles and the technical support provided by Barcelona Supercomputing Center (RES-FI-2020-2-0020). I.L. Kurbakov and Yu.E. Lozovik are supported by the Russian Foundation for Basic Research, Projects No. 19-02-00793 and 20-02-00410.

%

\end{document}